# The High Energy Telescope on *EXIST*

J. Hong[a*], J. E. Grindlay[a], B. Allen[a], S. D. Barthelmy[b], G. K. Skinner[b], N. Gehrels[b], and the *EXIST* HET Working Group

[a]Harvard Smithsonian Center for Astrophysics, 60 Garden St., Cambridge, MA 02138,
[b]NASA Goddard Space Flight Center, Greenbelt, MD 20771

## ABSTRACT

The Energetic X-ray Imaging Survey Telescope (*EXIST*) is a proposed next generation multi-wavelength survey mission. The primary instrument is a High Energy telescope (HET) that conducts the deepest survey for Gamma-ray Bursts (GRBs), obscured-accreting and dormant Supermassive Black Holes and Transients of all varieties for immediate followup studies by the two secondary instruments: a Soft X-ray Imager (SXI) and an Optical/Infrared Telescope (IRT). *EXIST* will explore the early Universe using high redshift GRBs as cosmic probes and survey black holes on all scales. The HET is a coded aperture telescope employing a large array of imaging CZT detectors (4.5 m$^2$, 0.6 mm pixel) and a hybrid Tungsten mask. We review the current HET concept which follows an intensive design revision by the HET imaging working group and the recent engineering studies in the Instrument and Mission Design Lab at the Goddard Space Flight Center. The HET will locate GRBs and transients quickly (<10–30 sec) and accurately (< 20") for rapid (< 1–3 min) onboard followup soft X-ray and optical/IR (0.3–2.2 μm) imaging and spectroscopy. The broad energy band (5–600 keV) and the wide field of view (~90º × 70º at 10% coding fraction) are optimal for capturing GRBs, obscured AGNs and rare transients. The continuous scan of the entire sky every 3 hours will establish a finely-sampled long-term history of many X-ray sources, opening up new possibilities for variability studies.

## 1. INTRODUCTION

A high redshift Gamma-ray Burst (GRB) (GRB090423) recently discovered by *Swift* and subsequently identified as being at *z*=8.3 by the followup observations from the ground is marked as a most distant object detected so far [1]. It demonstrated the presence of stars as early as 400 million years from the big bang and illustrated the high redshift GRBs as a rare tool to probe the early Universe. The Energetic X-ray Imaging Survey Telescope (*EXIST*) is a proposed multi-wavelength observatory to detect and identify these high redshift GRBs and use them to study the Epoch of Reionization (EOR). *EXIST* will discover black holes (BHs) on all scales and monitor the transient sky on unprecedented time scales. The mission concept for *EXIST* has been studied as a Black Hole Finder Probe under NASA's Beyond Einstein program, and the new concept is the result of an intensive design study under a NASA grant from the Astrophysics Strategic Mission Concept Study program, followed by the recent studies in the Instrument and Mission Design Labs (IDL and MDL) at GSFC.

The current concept of the *EXIST* mission consists of three instruments – the High Energy Telescope (HET), Soft X-ray Imager (SXI) and Optical/Infrared Telescope (IRT) – working together to maximize the science return. The HET, the flag ship instrument of *EXIST*, is redesigned to detect and locate the GRBs and transients rapidly while improving the original scientific objectives of surveying transient hard X-ray sky and also allowing room for two new longer wavelength instruments in the payload of an Evolved Expendable Launch Vehicle (EELV) (4 m fairing). The HET is a wide-field hard X-ray coded-aperture telescope covering 5 – 600 keV band with a ~ 90º × 70º field of view (FoV) using a large array of imaging CZT detectors and Tungsten coded aperture masks. Thanks to the design heritage from the past and current missions such as *Swift* [2] and the on-going technology development of upcoming missions such as *NuSTAR* [3], all the subsystems in *EXIST* enjoy a high maturity of technology, and it can be launched as early as 2017, provided the mission starts in 2012.

---

[*] Send correspondence to J. Hong (jaesub@head.cfa.harvard.edu). For easier reading, see the color version of this paper posted on http://arxiv.org/archive/astro-ph or at http://hea-www.harvard.edu/ProtoEXIST/.

In this paper, we review the *EXIST* mission (§2), and describe the new HET design (§3). We also outline our plan to achieve the new technology needed for HET (§4). Skinner et al. (2009) [4] describes the HET imaging and Allen et al. (2009) [5] the CZT detector plane development using our *ProtoEXIST* balloon program. For the details of the SXI, see Natalucci et al. (2009) [6] and Tagliaferri et al (2009) [7] and for the IRT, see Kutyrev et al. (2009) [8].

## 2. THE *EXIST* MISSION OVERVIEW

The *EXIST* mission concept consists of three complementary instruments as shown in Fig. 1. The HET covers 5 – 600 keV with a 90º×70º FoV, the SXI for 0.1 – 10 keV with 30′ FoV and the IRT for 0.3 – 2.2 μm with ~5′ FoV. The mission parameters are summarized in Table 1. The primary scientific objectives of the *EXIST* mission are to (1) Discover and identify high redshift GRBs and use them to probe the EOR (2) Discover the supermassive black holes (SMBH) either obscured or dormant to explore their evolution and influence in the cosmic landscape and thus to identify the total accretion luminosity of the Universe (3) Discover the rare X-ray transient events to reveal the nature of compact objects, their progenitors, and space-time. *EXIST* will explore black holes' birth (e.g. GRBs), growth (e.g. SMBHs) and influence in the cosmic landscape by discovering black holes on all scales from stellar mass to possible intermediate to supermassive ones in the center of the galaxies,

In the first two years of the mission, the HET FoV will scan nearly the full sky every two orbits with ±~25º offset to north or south in every other orbit, and the next three years, *EXIST* will follow up on half of the ~40,000 selected sources discovered during the first two years. Throughout the 5 years of the mission lifetime, *EXIST* will always immediately follow up on GRBs and bright transients for two or three orbits per event for optical/IR identifications and redshifts. *EXIST* will detect about 400 – 600 GRBs per year with >10–50 GRBs at $z > 8$. The continuous scan of the X-ray sky in the hard X-ray band will open a new timing domain of studying highly variable X-ray sources, and allow us to discover rare transient events including rapid bursts (< a few sec) to elusive tidally disrupted events of stars (~ weeks to months). During the scanning phase of the mission operation, the SXI will also survey roughly the half of the sky and the combined broad energy band coverage (0.1 – 600 keV) between the SXI and the HET provides a unique broad band spectrum of the X-ray emission mechanism and nature of the sources. For instance, for an AGN, this spectral coverage unambiguously identifies them as either obscured or unobscured or if obscured, as Compton-thick or thin.

For a given GRB or transient event, the HET will localize the source within 20″ in 10 sec, which will initiate the rapid slew of the whole observatory. The target will be acquired in the FoV of the SXI and IRT within <100 sec, and the SXI will provide the finer localization (<2″) within 10 – 100 sec, and the IRT will promptly identify the target by its variability and the fine source coordinates supplied by the SXI. The IRT will acquire on-board low (R=30) or high resolution (R=3000) spectra of the target within 300 - 2000 sec. This rapid response will identify the Ly-alpha break for the redshift measurements of GRFBs at $z \geq 2$, and high resolution spectroscopy allows detailed studies of the environment where the GRB occurred.

**Table 1:** *EXIST* mission parameters.

| Parameters | Values |
|---|---|
| Orbit | 600 km, 15º–22º inclination, 5yr mission |
| Mode | Zenith orbital scan (2yr); inertial pointing (3yr) |
| High Energy Telescope (HET) | 5–600 keV, 90º×70º, ≤20″ centroiding (90%) 0.08–0.4 mCrab (<150 keV, 1yr survey, 5σ) 0.5–1.5 mCrab (>150 keV, 1yr, 5σ) |
| Optical/IR telescope (IRT) | 0.3–2.2 μm, 4'×4', 0.3″ resolution AB~24 mag in 100 sec |
| Soft X-ray Imager (SXI) | 0.1–10 keV, 20'×20', 15″ resolution $2 \times 10^{-15}$ erg cm$^{-2}$ s$^{-1}$ (10 ks) |
| Spacecraft (S/C) | Pointing: 10″ transverse, 150″ roll Knowledge: 2″ (90% conf.) |
| Mass | 5497 kg (20% contingency) + 434 kg (propellant) |
| Power | 2803 W (30% contingency) |
| Telemetry | 90 GB/day |
| Launcher | EELV with 4m fairing (e.g. Atlas V-401) |

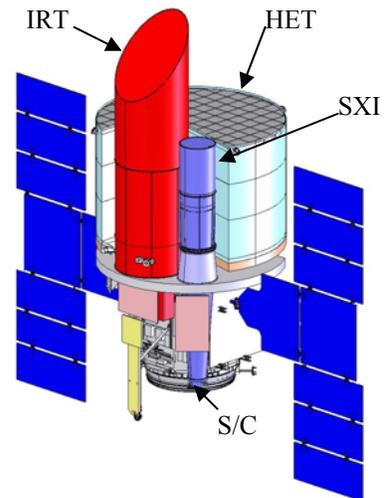

**Fig. 1:** The *EXIST* Observatory

Although *EXIST* shows some resemblance to the current GRB mission *Swift* (e.g. 3 instruments with rapid slew capability), all three of its instruments are ≥10X the sensitivity of their *Swift* counterparts. Some of the key changes are 1) inclusion of the IR band (both imaging and spectroscopy to identify high redshift GRBs (z>7), 2) capability of onboard optical/IR spectroscopy for rapid identification of fading events, (3) scanning mode operation in the first two years to maximize the sensitivity of the HET, and (4) broader energy band coverage with a finer angular resolution in the HET compared to *Swift*/BAT.

## 3. THE DESIGN CONCEPT OF THE HET

The HET is a wide-field hard X-ray coded-aperture telescope to make the initial detection and localization of GRBs and transients for the followup observations and conduct a sensitive hard X-ray survey. The hard X-ray band (5–600 keV) is ideal for detecting energetic GRBs, AGN, and transient sources because of the relatively low density of steady background sources, relatively low absorption by material surrounding sources and relatively high brightness of transients allowing precise localization. The HET has the longest design history of the three *EXIST* instruments. The current concept shown in Fig.2 is the result of extensive studies of various design configurations under the ASMC and the design was further refined during the IDL and MDL studies. The HET employs large arrays of fine pixel CZT detectors and a hybrid tungsten mask. Table 2 summarizes the key parameters of the HET in comparison with *Swift*/BAT. Like all coded-aperture instruments, the design is very tolerant to losses of detector pixels or modules. The HET has a rich design heritage from *Swift*/BAT and *INTEGRAL*/IBIS in both hardware and software. Its operational modes (e.g. scanning mode) benefit from the experience in *Fermi*/LAT. The detector technology greatly benefits from the upcoming mission *NuSTAR* in addition to our balloon-borne program *ProtoEXIST* (§6).

Fig. 3 shows the estimated 1-yr survey sensitivity of the HET in comparison with other missions and illustrates the 1 orbit sky coverage of the HET. The 1-year survey sensitivity of the HET is expected to be ~0.1–0.2 mCrab, depending on the energy range. The continuous scan with the wide field of view (~90º × 70º at 10% coding fraction) increases the chance of capturing rare elusive events such as soft Gamma-ray repeaters and tidal disruption events of stars by dormant SMBHs. Sweeping nearly the entire sky every two orbits (3 hours) will also establish a finely-sampled long-term history of the X-ray variability of many X-ray sources, opening up new possibilities of variability studies. Continuous scanning is also desirable for optimal imaging performance by averaging out unknown systematic noises (see §4.2). Full-sky images are updated onboard every ~1 s over a series of time intervals (2–1000 s) and energy bands and continually scanned for transient sources. After the initial 2 years of the full sky survey in scanning mode, *EXIST* will be primarily in pointing mode for followup studies: X-ray/Optical/IR spectra for source identification and redshifts of selected samples of ~20,000 AGNs in the survey. GRB detections and afterglow followups will continue throughout the mission lifetime (>5 years).

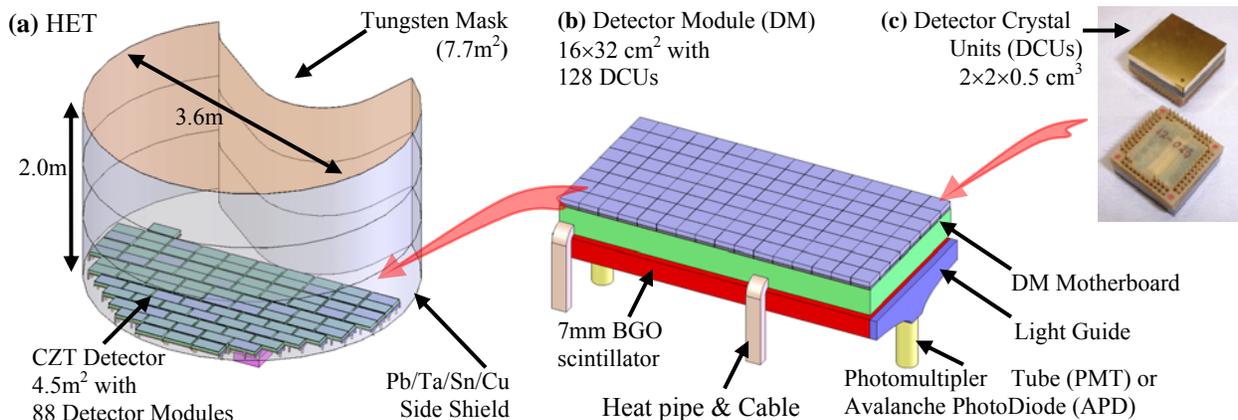

**Fig. 2 (a)** The HET design overview and the CZT detector plane consisting of **(b)** Detector Modules (DMs), which in turn consist of **(c)** Detector Crystal Units (DCUs).

**Table 2:** Key parameters of EXIST/HET vs. Swift/BAT

| Parameters | *EXIST*/HET | *Swift*/BAT |
|---|---|---|
| Telescope | 4.5m$^2$ CZT detector + 7.7m$^2$ w hybrid tungsten mask | 0.5m$^2$ CZT detector + 2.7m$^2$ Pb mask |
| Energy range | 5 – 600 keV (5mm thick CZT) 600 – 2000 keV (BGO) for GRBs | 15 – 200 keV (2mm thick CZT) |
| Sensitivity (5σ) | 0.1–0.4 mCrab (<150 keV, ~1 yr survey) 0.5–5 mCrab (>150 keV, ~1 yr survey) 20–40 mCrab (<150 keV, ~10 s on-axis) | 1 mCrab (<150 keV, ~2 yr survey) |
| Field of view | 90° × 70° (10%) | 110° × 90° (10% coding) |
| Angular & Positional resolution | 2.4′ resolution 20″ pos. for 5σ source (90% conf. Rad) | 17′ resolution 3′ pos. for 5σ source |
| Sky coverage | Nearly full sky every two orbits (3hr) | 10s orbits – a few days |
| Spectral Resolution (FWHM) | 2 – 3 keV (3% at 60 keV, 0.5% at 511 keV) | 3 – 4 keV (5% at 60 keV) |
| Timing resolution | 10 μsec | 100 μsec |
| **Components** | | |
| CZT detector | 2 × 2 × 0.5cm$^3$, 0.6mm pix. size, 12M pix. total, 11264 crystals | 4 × 4 × 2mm$^3$, 4mm Det. = pix. size, 32K pix. total, 32768 crystals (256 modules) |
| Mask | Tungsten Coarse elements: 15mm pixel, 3mm thick Fine elements: 1.25mm pixel, 0.3mm thick | Lead 5mm pixel, 2 mm thick |
| Side Shield | Pb/Ta/Sn/Cu | Pb/Ta/Sn/Cu |
| Rear shield | 7mm thick BGO Scintillators | Pb/Ta/Sn/Cu |

### 3.1 The Imaging and Mask Design

In order to meet the ambitious scientific goals of the *EXIST* mission which challenge conventional mask designs in coded-aperture telescopes, the HET uses a hybrid pattern consisting of two different pixel scales [9] as shown in Fig.4a. For instance, the wide energy coverage calls for a relatively thick mask (≥ 3mm even with Tungsten) and the fine angular resolution requires small pixels (~1.2mm). The high pixel to thickness ratio results in severe vignetting (auto-collimation) for off-axis X-rays, limiting the FoV, however a large FoV is required for the efficient detection of GRBs, transients, and for monitoring highly variable X-ray sources. The hybrid mask in the HET overcomes these difficulties in conventional coded-aperture masks by achieving simultaneously: (1) Fine angular resolution, (2) Wide energy band

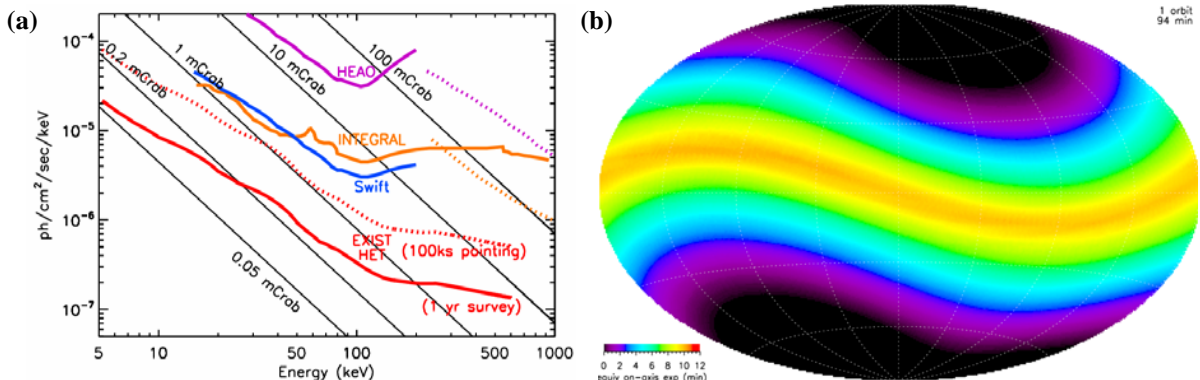

**Fig.3 (a)** 1-yr survey sensitivity (5σ) and **(b)** sky coverage of a 1 orbit scan in orbit plane. Full-sky coverage is obtained by alternatively scanning above/below the orbital plane by +/-25°.

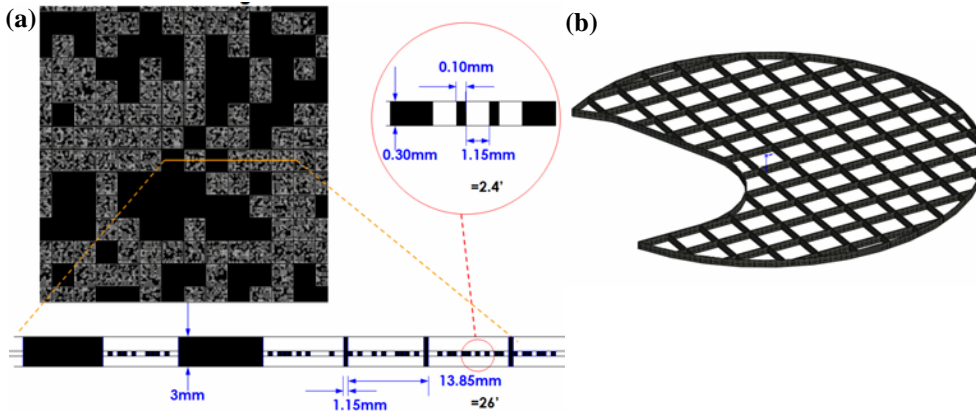

**Fig. 4 (a)** A segment of the HET hybrid mask with cross sectional view: coarse (15mm pitch, 3mm thick) and fine (1.25mm pitch, 0.3mm thick). **(b)** The grid structure that supports the tungsten Mask. The lowest vibration mode is at 35 Hz.

and wide field coverage, with minimal auto-collimation, and (3) Rapid and fine (<20") source localization.

The 3mm thick coarse elements (15mm pixel, 50% open fraction) are effective over the whole 5–600 keV energy band, allowing for the wide field imaging by minimizing off-axis losses due to vignetting. In addition, the hybrid mask enables the implementation of a two step imaging and localization processing, which dramatically reduces the image reconstruction computational requirements [4]. The fine elements (1.25mm pixels, 0.3 mm thick) are effective below ~100 keV. They are superimposed on the coarse mask, giving 25% overall open fraction below ~100 keV and ~50% above ~200 keV. The fine mask allows the precise location of events found in the rapid coarse-mode analysis. Because only small regions around candidate locations need to be imaged, this, too, is rapid. Despite the slightly reduced sensitivity of the coarse-mode analysis, the overall detection threshold is not impaired because all candidates are followed up in the fine mode to see if they are significant. The hybrid mask is built from stacked 0.3mm thick tungsten sheets with etched (low-cost, high-precision) mask holes.

The full mask consists of about 11 × 11 separate segments, each of which is about 30 × 30 cm$^2$. The cutout in otherwise a circular disk shape of the mask is to allow room for the SXI and IRT (see Fig. 1, Fig. 3a and Fig. 4b). The IRT baffle sticks out above the mask plane, but it is made of low-Z elements mostly transparent to X-rays, and thus the impact on imaging is minimal (although it does cast a faint gnomon-like shadow at low energies). The coarse mask in each segment consists of 10 stacked Tungsten sheets (0.3 mm thick each) where the open elements are chemically etched out. The fine mask consists of a single 0.3 mm Tungsten sheet imbedded in the center of the coarse mask. Chemical etching is a low cost high precision (the etching pattern error is <10% of the thickness, e.g. <300μm pattern error for a 0.3 mm thick sheet; the thickness is uniform within 1%) technique to generate the coded-pattern in thin Tungsten sheets. Prototype laminated masks Tungsten have been developed and fabricated for ProtoEXIST1 [5].

The mask support structure has been designed to minimize the obscuration of the detector plane, using a rectangular grid pattern composed of carbon fiber (Fig. 4b). Each segment of the mask sits on a matching grid block whose dimensions are 30.6 (length) × 0.635 (width) × 7.1 (vertical thickness) cm$^3$. This introduces moderate collimation effects for large off-axis X-ray photons. The whole structure is supported by a cylindrical carbon fiber support structure that is attached to the perimeter of the mask. The stress analysis on the HET structure shows that the first vibration mode is at 35 Hz or higher.

### 3.2 Detector Plane

As shown in Fig. 3a, the detector plane spans about 4.5 m$^2$ and consists of 88 identical detector modules. Each detector module consists of 128 Detector Crystal Units (DCUs). Each DCU is made of a 2×2×0.5 cm$^3$ CZT crystal (32×32 pixels, 0.6 mm pixel) bonded to an *EXIST*-specific ASIC (EX-ASIC), with a matching 2-D array of 32×32 channels. The CZT detector plane of the HET is hierarchically modular both in mechanical packaging and in the data concentration and processing. Based on the architecture implemented on *Swift*/BAT, the CZT detector allows both redundancy and fast imaging. To reduce background, the detectors are surrounded by graded-Z passive side shields (Pb/Ta/Sn/Cu) and Bismuth Germanate (BGO) rear anti-coincidence shields. The latter also extends GRB spectral measurement up to a few MeV.

CZT is the optimal choice for the detectors because of its combination of: (a) High-Z for gamma-ray detection efficiency, (b) Good position resolution (≤0.6mm) with the pixilated monolithic detectors, (c) Low cost, (d) Adequate energy resolution at room temperature. In addition, their use in space has been successfully demonstrated in *Swift*/BAT (and *INTEGRAL*/IBIS with CdTe detectors). The high maturity of CZT production driven by medical and Homeland security application delivers high quality CZT crystals at low cost. Currently there are three major CZT suppliers, Redlen technologies, eV products, and Orbotech Ltd., along with a few other firms that may begin large scale production in the near future. For instance, Redlen technologies, which have developed a traveling heater CZT growth technique, provides highly uniform CZT crystals at low cost. These crystals exhibit very low leakage current (<8 nA/cm$^2$) and excellent spectral performance: 3.2 keV resolution (FWHM) at 60 keV using 2.5mm pixels read out with the RadNet ASIC [10].

The combination of the large area (4.5m$^2$) and the fine pixels (0.6mm) of the CZT detectors required for rapid and precise localization of GRBs and transients demands a readout system with a large number of electronic channels or pixels (~11.5M). The power constraint (~240W) for the front-end electronics (FEE) requires low-power EX-ASICs (~20µW/pixel). It is not easy to increase the FEE's power budget since enlarging solar panels and the battery would be difficult due to fairing constraints. The larger sizes also would result in slower slew times due to higher moments of inertia. The primary challenge for the implementation of CZT detectors on the HET is the development of FEE that are capable of handling signals from a large number of pixels (11.5M) with limited resources, especially the power.

The EX-ASIC is essentially the direct bonded (DB)-ASIC developed for the CZT detectors in *NuSTAR* [3]; but with a modification for lower power. Except for power consumption, the DB-ASIC meets all requirements for the EX-ASIC, including the form factor (0.6mm pixel size), multi-pixel pulse-profile readout to enable depth sensing, wide dynamic range (1 – 1000 keV), and low electronics noise (<0.4 keV FWHM). The power consumption of the current DB-ASIC is already relatively low (~80µW/pixel with ~1 keV FWHM energy resolution and 0.3 keV FWHM electronics noise), but is four times higher than required. Because of the inverse relation between power and noise and the ~2 keV vs. <~0.4 keV electronics noise (FWHM) requirements for *EXIST* vs. *NuSTAR*, the modification for lower power is straightforward and will be developed under the balloon-borne experiment, *ProtoEXIST2*.

### 3.3 Passive Side Shield

We use a passive fringe shield to set the HET FoV by limiting the background low energy X-rays hitting the detector without passing through the mask. The fringe shield is composed of multiple layers of metal foil (Pb/Ta/Sn/Cu), which absorb incoming background X-rays or cascade them down to fluorescent X-rays at lower energies. The fringe shield is attached to the annulus mask support structure and spans from the detector plane to the mask plane. It is composed of three different sections with varying thickness (0.16/0.08/0.08/0.04 mm at top, 0.24/0.16/0.16/0.04 mm at middle, 0.36/0.24/0.24/0.8 mm at bottom near the detector plane). The thickness of each layer has been determined through Monte Carlo (MC) simulation in order to ensure that the shield efficiently stops X-rays at low energies (<200 keV), where the dominant background component is the diffuse cosmic X-ray background.

### 3.4 Onboard Imaging and Data Processing

To enable rapid followup observations of GRB afterglows, the HET needs to locate the GRBs quickly (<10 sec) and precisely (<20″). Since GRBs occur randomly in the sky and the prompt signals can last from a fraction of a second to ~1000 sec (for high redshift GRBs, it may thus take as long to detect the signal), the HET keeps and updates the full sky image every two sec in a set of energy bands (e.g. 5-15, 15-50, 50-150, 150-300, 300-600, 10-300 keV) and time intervals (e.g. 2, 8, 32, 64, 128, 512 sec). Although computationally challenging, the onboard imaging processing can be achieved with redundancy using fast processors that are becoming available and taking advantage of the substantial reduction of operations enabled with the two-step imaging process - the coarse Fast Fourier Transformation (FFT) image of the full sky followed by the local back-projection fine image around the candidate source which is made possible through use of the hybrid mask.

The onboard computers in the HET process the real time data for the fast GRB trigger and pass the full data stream to the spacecraft for downlink. The fast onboard imaging procedure has been designed and the onboard computers configured accordingly, based on our experience with *Swift*/BAT, in particular for the BAT Slew Survey (BATSS, [11]), which continuously processes the slew data of the BAT for GRB and transient searchs in real time on ground as they are downloaded with scheduled telemetry dumps. On *EXIST* this would done automatically on board.

As shown in Fig. 5, the detector plane is divided into 4 detector quadrants (DQs). Each DQ has its own dedicated processor box with 4 HyperX processors (~25 GFlops/HyperX) controlled by a LEON3FT processor. Each DQ generates coarse FFT sky images using the coarse mask pattern from the data of 22 DMs (520k operation, 9.5ms/image under the assumption of a conservative 18% efficiency for HyperX), and using just 2 HyperX processors with other 2 for backup. Then the main processor box, equipped with 2 Tileara processors (~50 GFlops/Tileara), collects and merges the coarse images from 4 DQs and searches for GRBs and transients. This processing repeats and updates a set of the full sky images every 2 sec. Once a GRB or transient is found (trigger), the main processor commands each DQ to generate a fine image of ~1×1 deg$^2$ around the source using a back-projection (BP) method with the fine pixel mask map (0.06sec/BP). Then, the fine images from 4 DQs are again collected and merged in the main processor, which allows for localization of sources with <20″ precision and within 10 sec of trigger. Note that for the fine localization there is no need to generate the fine images for all the sets of the energy bands and intervals but onlyh the fine image for the energy band and time interval with the strongest detection in the coarse images is needed. Finally the HET main processor alerts the spacecraft to initiate the slew based on the coordinates of the source and to downlink the fast GRB alert to the ground through the Tracking and Data Relay Satellite System (TRDSS). Meanwhile the raw event data continues to flow to the spacecraft for the full data downlink.

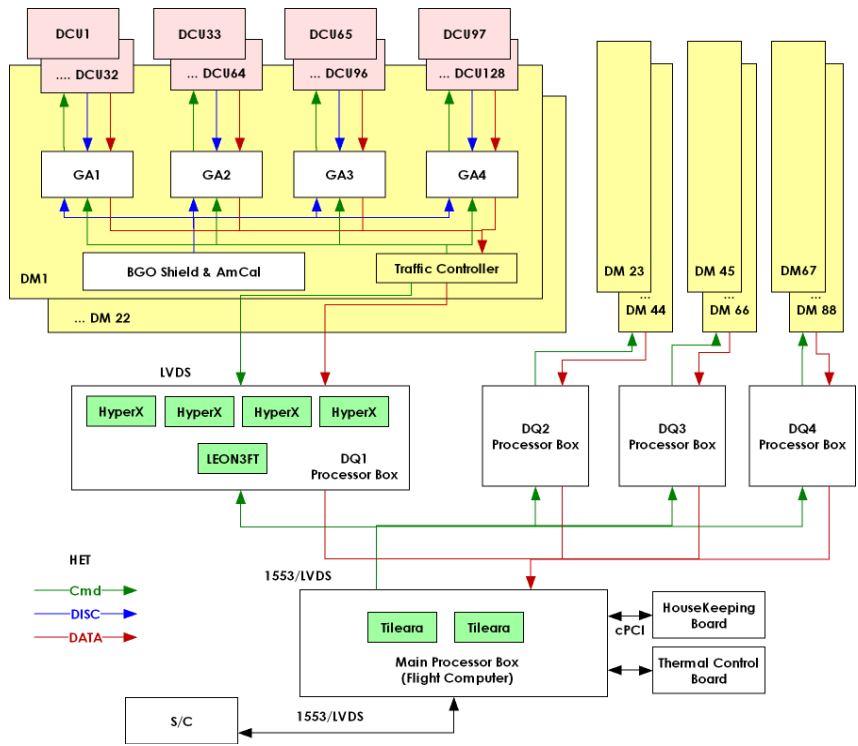

**Fig. 5** The HET backend electronics and data processing architecture. The CZT detector plane is divided into four Detector Quadrants (DQs), each of which is processed by a DQ processor box.

## 4. TECHNOLOGY DEVELOPEMENT

### 4.1 low power ASICs and CZT+ASIC hybridization

Table 3 summarizes the development road map of the FEE and the ASIC for *EXIST*/HET. A series of the balloon-borne experiments are scheduled to bring the ASIC and electronics packaging for large-array CZT detectors to the point of space qualification. The first flight, *ProtoEXIST1*, scheduled for fall 2009, will qualify the packaging concept using 64 CZT+RadNET ASIC hybrids (2.5mm pixels). The RadNET ASIC belongs to the same family of the ASIC development that resulted in the DB-ASIC. The main differences in the two are the form factor and the inclusion of the Analog-to-Digital Converter (ADC) in the DB-ASIC. The RadNET ASIC contains a 1-D array of 64 channels, connected through an interposer board to a 8 x 8 pixel (2.5mm) single CZT detector, whereas the DB-ASIC has a 2-D array of 32×32 pixels (0.6mm) that are direct-bonded to the 2D pixels on the ASIC.

**Table 3** Electronics noise (FWHM) and energy resolution (FWHM) of FEE ASIC development

|  | **Rev1 (1-050)** | *ProtoEXIST1* | *ProtoEXIST2* | *ProtoEXIST3 & EXIST* |
|---|---|---|---|---|
| **ASIC** | RadNET<br>70 μW/pixel<br>64 channels | RadNET<br>70 μW/pixel<br>64 channels | DB-ASIC**<br>80 μW/pixel<br>1024 channels | EX-ASIC<br>20 μW/pixel<br>1024 channels |
| **ASIC (+ IPB)** | 2.8 | 2.1 | 0.4 (No IPB) | 2.0* (No IPB?) |
| **+ Crystal** | 3.7 | 2.7 |  |  |
| **+ HV (–600V)** | 4.5 | 3.0 | 1.3 | 2.5* |
| **60 keV (241Am)** | 4.7 (7.9%) | 3.2 (5.4%) | 1.3 (2.2%) | 2.5 (3.5%)* |
| **122 keV (57Co)** | 5.5 (4.2%) | 3.7 (3.0%) | 1.5 (1.2%)* | 2.4 (2.0%)* |
| **356 keV (133Ba)** | 7.8 (2.2%) | 5.3 (1.5%)* | 2.2 (0.6%)* | 3.5 (1.0%)* |
| **662 keV (137Cs)** | 7.9 (1.2%) | 5.4 (0.8%)* | 2.3 (0.3%)* | 3.5 (0.6%)* |
| **TRL6** |  | Fall 2009<br>(*ProtoEXIST1*) | Fall 2010<br>(*ProtoEXIST2*) | Fall 2011<br>(*ProtoEXIST3*) |
| **References** |  |  |  |  |

*Estimates
**For *NuSTAR*

The readout system of the DB-ASIC benefited from the experience with the RadNET ASIC, but the core architecture of the readout system in both ASICs remain identical (e.g. 16 continuous sampling capacitors per pixel), which makes the transition from the RadNET ASIC to the DB- or EX- ASICs straightforward. The second flight, *ProtoEXIST2*, slated for fall 2010, will qualify the CZT+DB-ASIC hybrids (0.6 mm pixels). And the third flight, *ProtoEXIST3*, scheduled for fall 2011 will qualify *EXIST's* Engineering Models using the EX-ASIC (0.6mm pixels). For ease of packaging and minimal gaps between detector modules and detector crystals (Fig. 2), the bonding to the EX-ASIC may be done with an interposer board (as with the RADNET ASIC, but now 2D-2D coupling).

For the CZT+ASIC hybridization, there are multiple CZT-bonding technologies that already are qualified for space application (e.g. conductive epoxy bonding as for the CZT detectors in *Swift*/BAT). For ProtoEXIST1, we have developed a TLPS bonding scheme in collaboration with Creative Electron, Inc., which may be applicable to ProtoEXIST2-3. The gold-stud+epoxy bond for CZT+DB ASIC detectors developed for *NuSTAR* is close to being qualified for space and is directly applicable to the CZT+EX-ASIC. Given the number of units needed for *EXIST and the likely need for an aggressive assembly schedule,* a fast turn-around, high-yield bonding technology is important. Therefore, we shall demonstrate the mass-production capability of these bonding techniques (e.g. TLPS vs. gold-stud).

We have been pursuing two highly efficient bonding techniques. The first is low- temperature solder bonding with typical turnaround (~2 days), and the second is a simpler, low-cost bonding technique, called Transient Liquid Phase Sinter (TLPS, developed by Creative Electron, Inc.), with even faster turnaround (<1 day). In particular, the *repeatable* nature of the TLPS bond guarantees a very high yield. These bonding techniques have already made great progress since we have successfully assembled many CZT+RadNET ASIC hybrids using each (e.g. [10]). These detectors will constitute the CZT detector plane in the *ProtoEXIST1* balloon payload and all three bonding techniques will continue to be refined through the *ProtoEXIST2* and *3* experiments.

### 4.2 Scanning coded-aperutre imaging and BATSS

The dynamic range or sensitvity of coded-aperture imaging in a conventional stationary pointing mode observation is known to be limited by unknown systematics. Although the long term survey by *Swift*/BAT demonstrated that the coded-aperture telescope can achieve nearly Poisson limited sensitivity vs. Exposure time (i.e. inversely proportionally to the square root of the time [12], the minimum flux limit is still some ~30% greater than that expected from Poisson statistics. This sensitivity limit of the *Swift*/BAT survey using superimposed pointed observations is achieved thanks to the aggressive correction on the known systematics (e.g. removing bright sources; monitoring for noisy detectors; etc.). In coded-aperture imaging it has been suggested that one can achieve a near ideal sensitivity even without the need for the sophisticated corrections on the systemactis by operating the telescope in a continuous scanning mode instead of steady pointing [13]. Scanning mode in coded-aperutre imaging would be analogous to an extreme version of dithering operation usually performed in the focusing as well as pointed coded aperture telescopes (both *Swift*/BAT and *INTEGRAL*/IBIS employ slow dithering patterns). The data collected in the scanning mode contains a much larger combination of the detector pixels relative to the sky pixels, and any systematics in the detector tends to average out and disappear in the resulting images. Therefore, in the first two years of operations, *EXIST* will be in a continous scanning mode, only interrupted by follow-up observations of GRBs and transients, which are expected to occur about two per day.

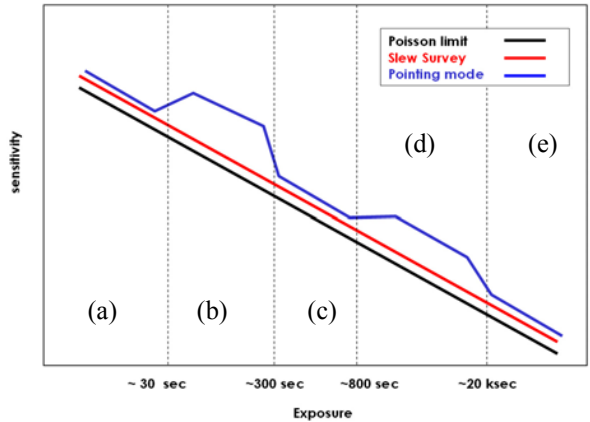

Fig. 6 Schematic overview of the sensitivity in both scanning (slew) and pointing modes. The difference between the slew and pointing modes in (a), (c) and (e) is not real and only for clarity. On average, the sensitivity of the slew survey (or the pointed observations in (a), (c) and (e) ) is about 10% higher than the Poisson limit.

In order to demonstrate the feasibility and performance of scanning coded-aperture imaging, the *Swift*/BAT slew survey (BATSS) is being carried out by collecting photon-mode data during scheduled slews (~60 per day) [11]. Fig. 6 shows a simplified version of the sensitivity as a function of time for both slew and pointed mode observations by *Swift*/BAT (for details, see [11] and [12]). For a short exposure (a), both modes are limited by low statistics, and thus their sensitivity is near Poisson limit. As the exposure increases (b), the systematics starts affecting the performance in pointing mode, and can only be corrected (partially) for exposures ≥300sec that are suffcient to establish the enough statistics for corrections on systematics (c). As the exposure further increases (d), unknown subtle systematics are left uncorrected by the analysis procedure affect the performance in the pointed mode. The superimposed data to achieve very long (~1-2Msec) exposures (e) consists of many separate pointings in various orientations, the residual systematics even in the pointed data is averaged out, resulting again near Poisson limited performancebut is still in excess (by ~30%) of that expected from Poisson statistics alone [12]. The BATSS data have not yet been extended into this very long exposure regime but are expected to be very close to the purely Poisson limit. Note the performance of the pointed data shown here (Fig. 6) is a represetive case and the actual performance can vary depending on the analysis. In the case of the scanning mode observations, thanks to the self-correcting nature of the data set, one can achieve very nearly Poisson-limited sensitivity without sophisticated correction procedure on the systematics. It is noteworthy that the scanning sensitivity is paricularly enhanced (vs. Pointing) for integrations ≤300 sec, so that the *EXIST* sensitivity for GRBs in particular is ~10X greater than that for *Swift*/BAT [14].

## 5. CONCLUSION

*EXIST* is a next generation multi-wavelength observatory, consisting of three telescopes: the HET, SXI and IRT. The HET is a wide-field coded-aerpture telescope, designed to capture rare eluive and exotic events such as high redshift GRBs in transient X-ray sky by monitoring the entire sky every three hours with high sensitivity and spatial and spectral resolution. The HET will employ advanced CZT imaging detectors with multi-scale Tungsten masks. The HET design benefits from the heritage of the past and current mission including *Swift*/BAT, and the detector technology takes advantage of the on-going development for upcoming mission *NuSTAR* in addition to the balloon borne prototype experiment *ProtoEXIST*. Scanning observations will allow the HET to achieve near-Poission limited performance.

# ACKNOWLEDGMENTS

This work is supported in part by NASA grants (for *ProtoEXIST*) NAG5-5396, NNG06WC12G, and NNX09AD76G and NNX08AK84G (for the Astrophysics Strategic Mission Concept Study for *EXIST*).

# REFERENCES

1. N. R. Tanvir et al. "A glimpse of the end of the dark ages: the gamma-ray burst of 23 April 2009 at redshift 8.3", submitted to Nature, arXiv:0906.1577, 2009.
2. N. Gehrels et al, 2004 ApJ, 611, 1005.
3. P. H. Mao et al, 2009, "Performance of the NuSTAR focal plane detectors" in this conference, *Proc. SPIE* **7435-2**
4. G. K. Skinner et al, 2009, "Imaging and burst location with the *EXIST* high-energy telescope" in this conference, *Proc. SPIE* **7435-10**
5. B. Allen et al, 2009, "ProtoEXIST1: wide-field hard x-ray telescope and initial prototype for *EXIST*'' in this conference, *Proc. SPIE* **7435-17**
6. L. Natalucci et al, 2009, "SXI on board EXIST: scientific performances *EXIST*'' in this conference, *Proc. SPIE* **7435-11**
7. G. Tagliaferri et al, 2009, "The X-ray mirror module of the Soft X-ray Imager (SXI) on board of the *EXIST* mission'' in this conference, *Proc. SPIE* **7437-5**
8. A. Kutyrev et al, 2009, "EXIST IRT imager-spectrometer'' in this conference, *Proc. SPIE* **7453-3**
9. G. K. Skinner and J. E. Grindlay, 1993 " Coded masks with two spatial scales", A&A, 276, 673
10. J. Hong et al 2009, "Building large area CZT imaging detectors for a wide-field hard X-ray telescope—*ProtoEXIST1*", Nuclear Instruments and Methods in Physics Research Section A, 605, 364.
11. A. Copete et al, 2009, in preparation.
12. J. Tueller, et al, 2009, "The 22-Month Swift-BAT All-Sky Hard X-ray Survey", submitted to ApJS, arXiv:0903.3037
13. J. E. Grindlay and J. Hong, 2004, "Optimizing wide-field coded aperture imaging: radial mask holes and scanning" in Optics for EUV, X-Ray, and Gamma-Ray Astronomy. Edited by Citterio, Oberto; O'Dell, Stephen L. *Proc. SPIE*, **5168**, 402.
14. J.E. Grindlay, "GRB Probes of the High-z Universe with EXIST," AIP Conference Series 1133, 18–24 (2009).